\documentclass[graybox]{svmult}

\usepackage{mathptmx}       
\usepackage{helvet}         
\usepackage{courier}        
\usepackage{type1cm}        
\usepackage{makeidx}         
\usepackage{graphicx}        
\usepackage{multicol}        
\usepackage[bottom]{footmisc}

\usepackage{listings} 
\usepackage{caption}

\makeindex             

\begin{document}

\title*{The Importance, Design and Implementation of a Middleware for Networked Control Systems}
\titlerunning{A Middleware for Networked Control Systems}
\author{Kyoung-Dae Kim and P. R. Kumar}
\authorrunning{K. D. Kim and P. R. Kumar}
\institute{Kyoung-Dae Kim \at Department of Electrical and Computer Engineering, University of Illinois at Urbana-Champaign, USA \email{kkim50@illinois.edu}
\and P.R. Kumar \at Department of Electrical and Computer Engineering, University of Illinois at Urbana-Champaign, USA  \email{prkumar@illinois.edu}}
%
%
\maketitle

\abstract*{Due to the advancement of computing and communication technology, networked control systems may soon become prevalent in many control applications. While the capability of employing the communication network in the control loop certainly provides many benefits, it also raises several challenges which need to be overcome to utilize the benefits. \\
In this chapter, we focus on one major challenge: a middleware framework that enables a networked control system to be implemented. Indeed our thesis is that a middleware for networked control sys important for the future of networked control systems. \\
We discuss the fundamental issues which need to be considered in the design and development of an appropriate middleware for networked control systems. We describe \emph{Etherware}, a middleware for networked control system which has been developed at the University of Illinois, as an example of such a middleware framework, to illustrate how these issues can be addressed in the design of a middleware. Using a networked inverted pendulum control system as an example, we demonstrate the powerful capabilities provided by Etherware for a networked control system.}

\abstract{Due to the advancement of computing and communication technology, networked control systems may soon become prevalent in many control applications. While the capability of employing the communication network in the control loop certainly provides many benefits, it also raises several challenges which need to be overcome to utilize the benefits. \\
In this chapter, we focus on one major challenge: a middleware framework that enables a networked control system to be implemented. Indeed our thesis is that a middleware for networked control sys important for the future of networked control systems. \\
We discuss the fundamental issues which need to be considered in the design and development of an appropriate middleware for networked control systems. We describe \emph{Etherware}, a middleware for networked control system which has been developed at the University of Illinois, as an example of such a middleware framework, to illustrate how these issues can be addressed in the design of a middleware. Using a networked inverted pendulum control system as an example, we demonstrate the powerful capabilities provided by Etherware for a networked control system.}


\section{Introduction}
\begin{figure}
\begin{center}
  \includegraphics[width=9cm]{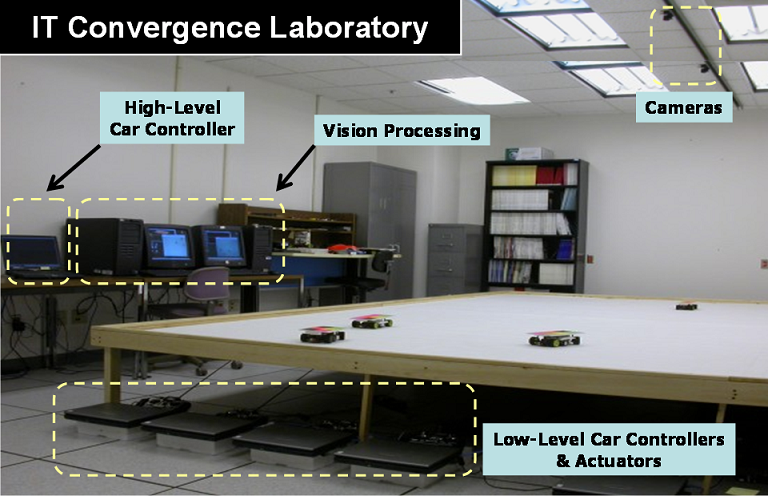}\\
  \caption{Traffic control Testbed in IT Convergence Laboratory at the University of Illinois}\label{fig:testbed}
\end{center}
\end{figure}

Over the past several decades, communication and computing technologies have advanced tremendously. Consequently, the platform for the control system itself also has changed with the emergence of networked control. In general, a networked control system is a system whose constituents such as sensors, actuators, and controllers are distributed over a network, and their corresponding control-loops are formed through a network layer. Thus, the scale of the networked control system is typically much larger than that of classical control systems. An example of such a system, an automatic traffic control system, established in the IT Convergence Laboratory at the University of Illinois, is shown in Fig. \ref{fig:testbed} (see \cite{convergencelab}).

In addition to the scale of the system, the complexity of a networked control system is also greater. Due to the existence of the networked communication and computation system below the control application layer, several challenging issues such as communication delay, the interface between a control application and the network layer, platform heterogeneity, clock differences between the computers, and others, arise. Clearly, all of these constitute an extraordinarily big burden on control engineers if they have to address these issues too, while designing of the control loops.

One solution to release these burdens from the control engineer is to interpose an abstraction layer between the application layer and the underlying networked communication and computation layer. Such an abstraction layer can encapsulate the complexity of the underlying system so that the application layer can have a much simpler view of the system. This can significantly simplify and shorten the development of a networked control application. Typically, such an abstraction can be realized as a software framework, called a \emph{middleware}. When such a middleware is designed and developed, it is important to consider the domain requirements which capture all the characteristics of the application domain. Thus, as a first step toward the development of the middleware for networked control systems, it is necessary to understand the fundamental characteristics of networked control systems and then establish corresponding requirements for the middleware framework.

\emph{Etherware} is such a middleware for networked control systems which has been developed at the University of Illinois \cite{baliga:05, kdkim:08}. In the sequel, we extensively discuss how Etherware is designed and how it works to support the domain requirements established from the domain characteristics, since it can serve as an exemplar of middleware for networked control systems. We also present a particularly demanding application that we have implemented, a networked inverted pendulum control system, as a case study of a particular networked control system which is implemented on top of Etherware, in order to demonstrate the usefulness of a middleware framework.

\section{Networked Control Systems} \label{sec:dcs}
\subsection{Domain Characteristics}  \label{sec:dcs:char}
There are many potential examples of networked control systems in various application areas, such as smart power grids, intelligent traffic control systems, automatic warehouse management systems, and so forth. In this section, we investigate the common characteristics which are shared by many networked control systems in many application domains.

\subsubsection{Large-scale}
In a networked control system, the control loop is typically formed through the underlying communication network. Thus, the physical distance between the entities in the loop is not an issue anymore. Also, the communication network allows us to form multiple control loops through it so that multiple control objectives can be achieved at the same time. The Testbed example shown in Fig. \ref{fig:testbed} is a good example which has such characteristics. In the testbed, a vision system is used to detect the state of the moving vehicle, and a low-level controller controls the vehicle to follow a given trajectory generated by a high-level controller. The inner control loop for tracking the given reference trajectory is formed through a communication network since all these elements are running at different computing nodes. Besides this inner control loop, there is another control loop formed through the same communication network to achieve a different slightly higher level control objective, which is collision avoidance between vehicles. In addition to all these, we also have another control loop in the testbed for runtime system management, such as upgrading or migrating some software modules to optimize the overall system performance.

\subsubsection{Openness}
In a classical control system, it is typically not allowed to change the running system. However, networked control systems are open to runtime system reconfiguration. While the system is running, new entities can join or leave the system, and also an existing entity can be replaced or even migrated to other location in a networked control system. Depending on the situation, the information flow which forms a control loop can be dynamically changed at runtime as well. The testbed system in Fig. \ref{fig:testbed} can exhibit all of these dynamic reconfiguration features. Clearly, vehicles can join or leave the traffic system dynamically. If a better traffic control algorithm is developed, then the existing traffic controller can be replaced by the new one with a better algorithm, so that the overall traffic control performance can be improved. Also, depending on the network traffic, a traffic controller may need to be migrated to another computing node to provide better traffic control performance. Once a controller is migrated, then the corresponding control loop also has to be reconfigured accordingly.

\subsubsection{Time-criticality}
A delay in a control loop typically affects the performance and the stability of the control system. Therefore, a control system is in fact a time-critical system in most cases. In fact, time-criticality is one of the fundamental characteristics of any control system, not just for a networked control system. Considering the fact that the computation and communication network are in the middle of control loop, the time-criticality might an even be more challenging issue for a networked control system.

\subsubsection{Safety-criticality}
In many cases, a control system is indeed a safety-critical system which can cause severe consequences once the system fails. As shown in the testbed, a vehicle control system can be an example of such a safety-critical control system. In a networked control system, it becomes is harder to achieve the safety-guarantee due to the existence of the computation and communication network.

\subsection{Domain Requirements} \label{sec:dcs:req}
The following are some of the requirements for a middleware framework for networked control systems.

\subsubsection{Operational Requirements}
As we discussed in Section \ref{sec:dcs:char}, entities which constitute a networked control system typically run on different computing nodes over the network. This distributed nature of a control system in fact causes several issues which have to be resolved for correct operation of a control system. The existence of the network itself can cause several difficulties in the development of a networked control system. Among them, the \emph{location difference} and \emph{clock difference} between entities in the control loop are two essential issues caused by the distributed nature. Thus, as an underlying platform of a networked control system, a middleware framework is necessary to provide an abstraction about the networked system which hides all such issues, so that a networked control system can be easily developed by the control designer. Besides these two requirements, it is also required to provide a mechanism which supports semantic addressing (or context-aware addressing) so that the portability and reusability of the application code can be enhanced.

\subsubsection{Management Requirements}
As explained in Section \ref{sec:dcs:char}, a networked control system is typically subject to runtime system reconfiguration since it is an open system. Even though it is still possible to implement all such functionalities in an application layer, it becomes much easier and more efficient to develop and manage a networked control system if the underlying platform is equipped with some functionalities which can be used for runtime reconfiguration. Thus, as a platform for a networked control system, a middleware framework is required to provide some mechanisms for \emph{runtime system management} which enables continuous system evolution.

\subsubsection{Non-functional Requirements}
The non-functional requirements\footnote{It should be noted that the phrase ``non-functional requirements'' can be used in dramatically opposite ways in different communities: with respect to the middleware designer, both a naming service or communication mechanism are both functional requirements, but achieving control loop stability is a non-functional requirement. From the viewpoint of the control designer, the reverse is true. In this paper, the viewpoint is that of the middleware designer.} for a middleware framework are induced from both the time-critical and safety-critical characteristics of a networked control system. The time-criticality requires a control system to behave in a predictably timely manner so as to minimize the effect of delay. Thus, a middleware framework is required to provide some mechanisms which support the \emph{timeliness} behavior of the control system. Also, the safety-criticality of a control system requires that the middleware framework itself be error-free, and also provide some mechanisms to tolerate faults which can occur in the application layer, to achieve overall \emph{reliability}.

\section{Middleware for Networked Control Systems} \label{sec:middleware}

\subsection{Middleware Fundamentals} \label{sec:middleware:fundamental}
A middleware is a software framework running in between an application and the underlying platform such as an operating system. Even a control system application running on a single computer can benefit from a middleware framework. However, the true value of a middleware is for a system which involves the features of both \emph{heterogeneity} and \emph{distributed operation}. Since these two fundamental features of distributed systems add a lot more complexity to the application, it is much harder to develop an application in general. Therefore, it is important to have a simpler abstract model of the system which hides all the complex details of the underlying system from an application developer. A middleware framework can provide such an abstraction of the system to the application developer so that she can develop an application easily on top of the abstraction. In this way, a distributed application can be developed more rapidly and reliably. In addition to rapid application development, an application can also be made more reliable since many of the commonly used functionalities which usually require expertise to handle low level complexities can be developed and provided to the application developer by a middleware as a form of middleware service. An example is \emph{component reuse} which can lead to a component economy.

\subsubsection{Communication Mechanisms}
In a networked control application, the distribution of the control system application over different nodes requires the interaction among components to occur over a network. For network programming, application programming interfaces such as \emph{sockets} are provided by an operating system. But these are usually low level and require some expertise to use. Also, they are tightly coupled to the underlying computer platform. Thus, they are not appropriate to be used in a platform independent way for developing a distributed application in general.

In contrast, a middleware framework can provide simpler \emph{inter-application communication mechanisms} to the application layer by encapsulating these low level network programming interfaces. A distributed application can then easily be developed using the inter-application communication mechanisms provided by the middleware, which allows components to interact with each other over a network without worrying about the low level network programming which is typically tedious and error prone.

In provisioning such communication mechanisms, there are roughly two forms of mechanisms which can be provided by a middleware, \emph{message-oriented communication} and \emph{request-oriented communication} \cite{puder:05}. In message-oriented communication, the message sender transmits a message to the receiver but the receiver does not respond to the sender. In contrast, the receiver replies with a response message when it receives a message from the sender in the request-oriented communication. Thus, the message-oriented communication can be considered as a one-directional communication model while the request-oriented communication can be considered as bi-directional communication.
Each of these communication mechanisms can be further classified as \emph{synchronous communication} or \emph{asynchronous communication}. The sender is passive (i.e., its execution is blocked) in synchronous communication, while it is active during the communication process in asynchronous communication. Considering the characteristics of a distributed system, the communication mechanism provided by a middleware is most fundamental to providing an abstraction of the original distributed system that eliminates several issues related to its distributed nature.

\subsubsection{Naming Service}
The other useful functionality which can be provided by a middleware framework is a \emph{naming service} which allows an application to easily find or communicate with other applications in a distributed system. It is still possible to develop a distributed application without having such a naming service; however it would require the explicit specification of the physical network address in the source code of the applications. A naming service provided by a middleware can eliminate this otherwise undesirable necessity. In fact, from the software engineering point of view, a middleware's naming service improves significantly the portability and reusability of the source code by breaking the connection between the application code and the underlying platform.

\subsubsection{Other Domain Specific Services}
Besides the above inter-application communication mechanisms and naming service, there are many other functionalities which can be provided as middleware services, such as security service, transaction service, event service, and so on \cite{vinoski:04}.

\subsection{Etherware} \label{sec:ether}
In this section, we continue our discussion about the middleware framework for networked control systems in the context of a specific middleware framework, called \emph{Etherware} \cite{baliga:05}, which has been developed at the University of Illinois.

\subsubsection{Domainware for Networked Control Systems} \label{sec:ether:4ncs}
Etherware is a middleware framework developed specifically for the networked control application domain. The main objective of Etherware is to provide a software framework which enables a rapid, reliable, and evolvable networked control application development. A networked control application can be easily developed in Etherware since it supports \emph{component-based application development}. A software component is a software module which provides a set of functionalities through a set of pre-defined interfaces. In Etherware-based applications, the pre-defined interface is used for interaction between a component and Etherware, and components interact with each other through Etherware. Thus, Etherware itself provides a virtual communication layer to the application layer. Etherware uses the message exchange mechanism for component interaction. More specifically, a component needs to create a message and sends it to Etherware. Then Etherware delivers the sent message to the receiver though its message delivering mechanisms. In Etherware, every message is an instance of \emph{Message} class which is a well-defined XML document object \cite{xml:w3c}. Listing \ref{lst:xml} shows the XML structure of the Message class.
\lstset{language=XML, caption=XML structure of an Etherware Message, label=lst:xml}
\begin{center}
\begin{lstlisting}
<EtherMsg type=... rel=... >
  <profile name=... ></profile>
  <content> ... </content>
  <ts value=... ></ts>
</EtherMsg>
\end{lstlisting}
\end{center}

The \verb|type| attribute of the \verb|EtherMsg| element is used to specify the name of the message. The name of the receiver component can be specified in the \verb|name| attribute of \verb|profile| element. In the \verb|content| element, any information concerning the interaction semantics can be specified. The clock time when the message is created is specified in the \verb|value| attribute of \verb|ts| element.

\subsubsection{Architecture} \label{sec:ether:arch}
\begin{figure} [htpb]
\begin{center}
  \includegraphics[width=9cm]{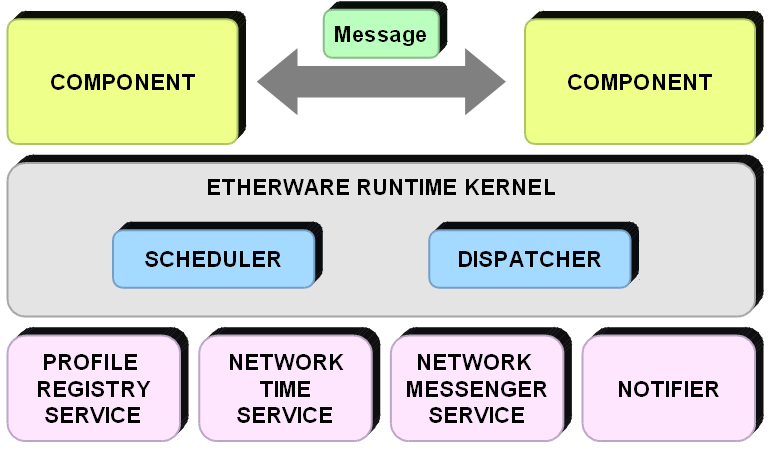}\\
  \caption{Etherware architecture}\label{fig:architecture}
\end{center}
\end{figure}

The concept of microkernel architecture in an operating system \cite{tan:07} is used in the design of Etherware architecture. Therefore, only the minimal invariant functionalities are implemented in the core module of Etherware, called \emph{Etherware Kernel}, while all the other functionalities are implemented as \emph{Etherware Components}. As shown in Fig. \ref{fig:architecture}, Etherware components can be classified further into two different layers. The top layer contains components which implement the application logic, called \emph{application components}, while the bottom layer is for components which provide functionalities to support several fundamental domain requirements, called \emph{service components}. The details of these components are explained in the following sections.

In this section, we discuss the Etherware Kernel which is a runtime platform for Etherware components. As a platform for component execution, Etherware Kernel is responsible for both component life-cycle management and message delivery between components. To deliver a message from one component to another, Kernel uses an object, called \emph{job}, as its scheduling entity. When a new message arrives in the Kernel, Kernel creates a new job which contains the sent message itself and the address of the recipient component. Then the Etherware Scheduler enqueues the job into a job queue. The enqueued jobs are processed one by one by a job processing software module, called \emph{Dispatcher}. When a Dispatcher processes a job, it first extracts the address information from the job and then it delivers the encapsulated message to the receiver component. During this delivery process, some of the service components can be called by Dispatcher depending on the delivery information specified in the message.

\subsubsection{Component Model} \label{sec:ether:component}
\begin{figure} [htpb]
\begin{center}
  \includegraphics[width=9cm]{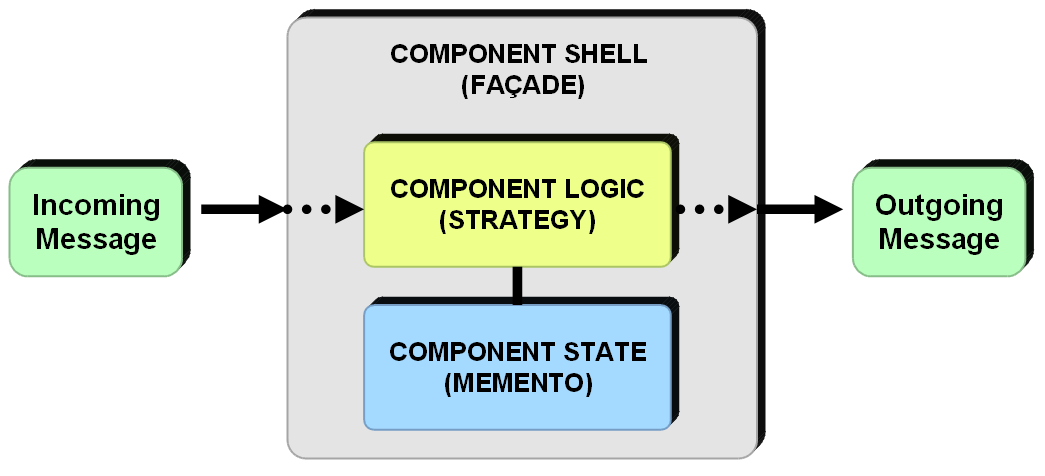}\\
  \caption{Etherware component model}\label{fig:component}
\end{center}
\end{figure}

Etherware's component model shown in Fig. \ref{fig:component} provides the framework in which an application component can easily be developed. In designing the component model, several \emph{software design patterns} \cite{GoF:94} were used. \emph{Shell} plays a central role in the component model. First, it manages the life cycle of a component which is encapsulated by it. Shell creates or destroys an instance of component as needed. Second, Shell in a component model provides a channel which allows a component to interact with other components.

Basic design to implement Shell is based on the \emph{Facade} design pattern. The \emph{Strategy} design pattern was used to design the component interface which defines a uniform interface between Shell and all components. Due to this Strategy design pattern, Shell can do runtime \emph{component replacement} since every component implements the same interface called \emph{Component Interface}. To support runtime component migration, another design pattern called \emph{Memento} was used in the component model. For \emph{component migration}, it is not enough to move a component by stopping at one place and by restarting at other place. To reduce the performance degradation due to component migration, the execution state of a component should be continued smoothly before and after the migration. The Memento design pattern was adopted to support this feature.

\subsubsection{Services} \label{sec:ether:svcs}
Etherware supports several fundamental functionalities which are commonly required for networked control system applications as Etherware services. \emph{ProfileRegistry} is a naming service which is implemented in Etherware to support the semantic addressing requirement. It maintains information about the profile of a component and its network address. \emph{NetworkMessenger} provides the service of message delivery over the network. It encapsulates all the details about network information such as protocol and network address. So, NetworkMessenger is the Etherware service which supports the domain requirement of hiding location discrepancy. NetworkMessenger is called only when a message is destined for a remote component since Etherware Kernel delivers the message directly if it is for local component. Etherware resolves the time discrepancy issue of distributed systems by implementing the \emph{NetworkTime} service. NetworkTime service translates a time stamp from the clock of a remote computing node to that of a local machine for every message which is received by NetworkMessenger. For this purpose, NetworkTime service maintains the clock offset and skew for every other computing node where an other NetworkTime service is running, by periodically exchanging \emph{Ping} and \emph{Response} messages. The \emph{Notifier} service provides a time-triggered message service to Etherware. Basically, Etherware is an \emph{event-driven system} such that a component gets executed only when it receives a message. However, in many cases, control actions need to be performed based on the \emph{time}. In such situations, the Notifier service enables a component to execute at the time that it has to, by sending a notification message to that component.

\section{Real-Time for Networked Control Systems}
Now we turn to the issues raised by the real-time operation that is necessary for control systems.

\subsection{Real-Time System Fundamentals} \label{sec:rt:fundamental}
A real-time system is a system whose correctness depends not only on the logical result of the computation but also on the \emph{time} at which the results are produced \cite{stankovic:88}. Thus, timeliness is a critical attribute of any real-time task. More specifically, the main objective of a real-time system is to meet the timing requirements of each real-time task. However, this does not necessarily mean that a real-time system is a fast system. Fast computing may \emph{help} in meeting timing requirements. However, fast computing alone does not \emph{guarantee} meeting timing requirements. In fact, a real-time system is more precisely a system which is predictable, whether fast or slow \cite{stankovic:88,buttazzo:04}. In general, \emph{predictability} is considered to be one of the most fundamental attributes of any real-time system. What one basically requires is that the system should behave in such a way that the execution behavior of the running task set can be precisely described from the information about both the system itself and the task set.

Once a system becomes predictable, then it is possible to address the issue of schedulability of a given real-time \emph{task set}. Each \emph{task} in the task set can be regarded as a collection of \emph{jobs}. Each job may have a \emph{deadline} (also called absolute deadline), which is the time by which the job must be completed. A task is said to have met its timing requirements if all the jobs in the task are completed by their deadlines. In a real-time system, a set of real-time tasks is said to be \emph{schedulable} if all tasks in the set can complete their execution, while satisfying all their timing requirements, under some task scheduling algorithm (or scheduling policy). Otherwise, the task set is said to be \emph{unschedulable}. A scheduling policy is a set of rules which determine the execution order, called \emph{schedule}, among the tasks in a given task set. However, determining schedules in a scheduling problem consisting of a set of $n$ jobs $J = \{J_{1}, J_{2},..., J_{n}\}$, a set of $m$ processors $P = \{P_{1}, P_{2},..., P_{m}\}$, and a set of $r$ types of resources $R = \{R_{1}, R_{2},..., R_{r}\}$, is known to be NP-complete \cite{buttazzo:04}. Therefore, it is important to consider a scheduling problem under additional assumption so that the problem becomes simple enough with respect to computational tractability while still preserving its practicality. Later, in this section, we introduce some of these fundamental real-time scheduling theories\footnote{For more details and comprehensive coverage of real-time scheduling, we refer the reader to \cite{buttazzo:04,sha:04}.}. Before discussing real-time scheduling, we first highlight the difference between a \emph{hard} real-time system and a \emph{soft} real-time system.

\subsubsection{Hard Real-Time vs. Soft Real-Time}
A real-time task has timing constraints which have to be met for its correctness. As noted earlier, one important time constraint is the \emph{deadline} by which a job in a real-time task has to complete its execution. Depending on the severity of the consequences which could occur by failure to meet the deadline constraints, real-time tasks are usually classified into two groups, \emph{hard real-time} tasks and \emph{soft real-time} tasks. A hard real-time task is a real-time task whose deadline constraints are very strict. The consequence of a deadline miss of a job in a hard real-time task could be catastrophic. In fact, many control tasks have the characteristics of a hard real-time task. One can easily imagine that an aircraft flight control task which has the highest priority among the tasks in any avionics system is indeed a hard real-time task. In contrast, the deadline constraints of a soft real-time task are not as strict as that of a hard real-time task. In a soft real-time system, an occasional miss of a deadline does not cause a catastrophic situation. However, the performance of the task might be affected by the rate or frequency of deadline misses. On-line multimedia streaming is a typical example of a soft real-time task.

\subsubsection{Processor Utilization Bound}
In many control systems, the sensing and control actions are typically periodic actions. Thus, a periodic task model in real-time scheduling theory can be used to capture the fundamental behavioral characteristics of control systems. In real-time scheduling theory, a periodic task can be described by several parameters such as the \emph{execution time} ($C$), the \emph{period} ($T$), and the \emph{relative deadline} ($D$). The \emph{period} is that length of the time interval between successive arrivals of jobs in the task. The \emph{relative deadline} is the time interval from the arrival of a job to its deadline, which is the latest time by which jobs must be completed. The \emph{execution time} is the amount of the processor's time that the job needs in order to be completed. For simplicity in schedulability analysis, it is usually assumed that the relative deadline of a job is the same as the period. With this assumption, the demand for the processor's time of a task set consisting of $n$ periodic tasks can be characterized by a parameter $U$, called \emph{processor utilization factor}, which is defined as follows:

\begin{equation} \label{eq:rt:util}
 U = \sum_{i=1}^{n} \frac{C_{i}}{T_{i}} .
\end{equation}

For a given scheduling policy and a given periodic task set, there exists a number $U_{ub}$, such that the schedulability of the task set is guaranteed if the process utilization factor of the task set is below $U_{ub}$. Above this upper bound, the given task set maybe unschedulable. Note that the value of $U_{ub}$ depends not only on the scheduling policy but also on the task set. A critical quantity is the least upper bound $U_{lub}$, defined as the minimum among all $U_{ub}$ over all task sets. $U_{lub}$ can be used the threshold so that it provides a sufficiency condition to test the schedulability of any task set under a specific scheduling policy.

\subsubsection{Rate Monotonic Scheduling}
Rate monotonic (RM) scheduling is central to several fundamental results in real-time scheduling. The rate of a task is defined as the reciprocal of its period. In RM scheduling, the execution priorities are statically assigned to tasks based on the rate of each task in a given set of periodic tasks. The higher the rate (i.e., the shorter the period), the higher the priority. RM scheduling is known to be optimal among all fixed-priority (or static) scheduling policies for periodic task sets \cite{liulayland:73}. It is optimal in the sense that if any other fixed-priority scheduling policy can schedule a given periodic task set, then the RM scheduling can also schedule the task set. In other words, there is no other fixed-priority scheduling policy which can schedule a given task set that is not schedulable under RM scheduling. In \cite{liulayland:73}, the \emph{least} upper bound of the processor utilization which guarantees schedulability of a periodic task set consisting of $n$ tasks is shown to be.

\begin{equation} \label{eq:rms:lub1}
 U_{lub} = n(2^{1/n} - 1) .
\end{equation}

If we take the limit as $n \rightarrow +\infty$ in (\ref{eq:rms:lub2}), we obtain $U_{lub}$ of RM scheduling for any periodic task set with any number of tasks in it:

\begin{equation} \label{eq:rms:lub2}
 \lim_{n \rightarrow \infty} U_{lub} = \ln 2 \simeq 0.69 .
\end{equation}

By using this condition, it is very easy to check the schedulability of a given task set under RM scheduling. If the processor utilization demand is below $U_{lub}$, then the task set is schedulable. Otherwise, there is no guarantee for a given task set to be schedulable. In this case, the task set can be either schedulable or unschedulable. To determine the schedulability of such task sets, it is necessary to employ iterative response time analysis which is a task set dependent, schedulability analysis technique \cite{buttazzo:04}.

\subsubsection{Earliest Deadline First Scheduling}
While RM scheduling is an optimal \emph{static} real-time scheduling policy, the \emph{earliest deadline first} (\emph{EDF}) is an optimal \emph{dynamic} real-time scheduling policy. In RM scheduling, a priority is assigned to a task and the assigned priority is never changed during the task's periodic execution. However, in EDF scheduling, a priority is assigned to each instance of a task, i.e., to each \emph{job}, based on the current job's absolute deadline. Therefore, a priority assigned to a task keeps changing depending on its current execution state. In \cite{liulayland:73}, the necessary and sufficient conditions for the schedulability of a set of periodic tasks under the EDF scheduling is shown to be :

\begin{equation} \label{eq:edf:lub}
 U = \sum_{i=1}^{n} \frac{C_{i}}{T_{i}} \le 1 = U_{lub} .
\end{equation}
where $C_{i}$ is the processor time necessary for executing the task $i$ without interruption and $T_{i}$ is the period of task $i$. Note that the optimality of EDF is immediate from (\ref{eq:edf:lub}) since the least utilization upper bound it provides for EDF scheduling is 100 percent of the processor time.

\subsubsection{Resource Sharing Protocol}
\begin{figure}
\begin{center}
  \includegraphics[width=9cm]{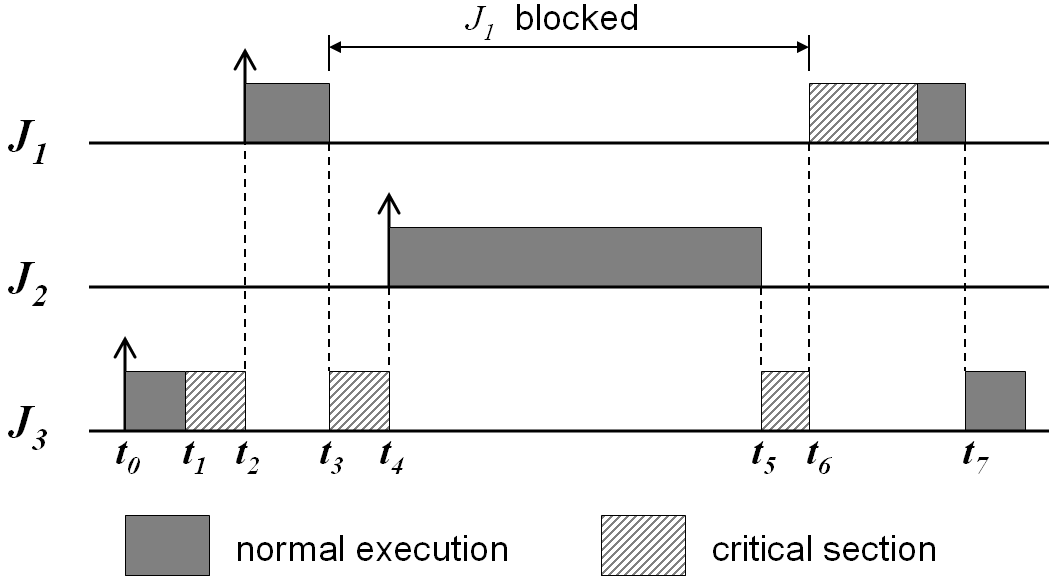}\\
  \caption{An example of priority inversion}\label{fig:rt:inversion}
\end{center}
\end{figure}

During its execution, a real-time task may access many resources such as data structures, files, memory areas, peripheral devices, a set of variables and so on. Also, in any real-time computing system, multiple tasks running concurrently may access the same resource. In some cases, accesses of the resource have to be mutually exclusive for the integrity of the sate of the shared resource at the same time. For this purpose, operating systems which manage the resources provide synchronization mechanisms to allow mutually exclusive access to the shared resource. To synchronize concurrent access from multiple tasks, the synchronization mechanism blocks a task when it tries to access a resource which is already occupied by another task, until it is released by the latter task.
However, synchronized resource sharing causes an undesirable phenomenon in real-time systems. In real-time systems, a task with the highest priority should be able to continue its execution under any circumstances. However, this may not be true any more under synchronized resource sharing. A higher priority task can be blocked by a lower priority task. This is called a \emph{priority inversion}. Fig. \ref{fig:rt:inversion} illustrates an example of priority inversion \cite{buttazzo:04}. In this example, $J_{1}$ has the highest priority, while $J_{2}$ and $J_{3}$ have intermediate and lowest priority, respectively.

In general, the duration of priority inversion can be potentially unbounded since any intermediate priority task such as $J_{2}$ in Fig. \ref{fig:rt:inversion} could indirectly block the highest priority. This in turn means that a real-time task can fail to meet its timing constraints due to priority inversion. To overcome this issue, a real-time resource sharing protocol, called \emph{priority inheritance protocol} (\emph{PIP}), was proposed in \cite{sha:90}. The basic idea of the PIP is that a task which currently holds the shared resource inherits temporarily the highest priority among the blocked tasks, until it releases the resource. After releasing the resource, it recovers its original priority. In this manner, the blocking task will never be preempted by any intermediate priority task while it is accessing a resource.
By adding PIP for resource sharing among real-time tasks in a given task set, the original RM schedulability test was extended in \cite{sha:90}. Any set of $n$ periodic tasks using the priority inheritance Protocol can be scheduled by the rate monotonic algorithm if the following conditions are satisfied:

\begin{equation} \label{eq:rms:pip}
 \forall i, ~1 \le i \le n, ~~~\sum_{k=1}^{i} \frac{C_{k}}{T_{k}} + \frac{B_{i}}{T_{i}} \le i(2^{1/i} - 1) .
\end{equation}
where $B_{i}$ is the maximum blocking time, due to lower-priority tasks, that a task $J_{i}$ may experience. $C$ and $T$ are the execution time and period of a task, as explained above.

\subsection{Real-Time Support in Etherware} \label{sec:rt:ether}
In this section, we describe how Etherware supports the real-time requirements of a middleware for networked control systems.

\subsubsection{Quality of Service (QoS) of Message Delivery}
In Etherware, each Etherware \emph{component} can be considered as a task in a real-time scheduling model. Also, each message sent by a component can be thought of as an instance of the task, a \emph{job} in the real-time scheduling model. As mentioned in Section \ref{sec:rt:fundamental}, each real-time task has a set of timing constraints such as period and relative deadline which are required to be met. In Etherware, these timing constraints can be specified as a QoS specification in each message sent by a component \cite{kdkim:08}. Such information is encoded as a QoS element in the Message class object which itself is an well-defined XML document, as explained in Section \ref{sec:ether:4ncs}.
Basically, Etherware defines QoS as a collection of attributes of an application which are used in scheduling for execution. Hence, any information which affects the scheduling can be specified as a constraint in QoS specification. In our current implementation of Etherware, the period (\verb|period|), the relative deadline (simply called \verb|deadline|), the worst case execution time (\verb|wcet|), and the importance of a message (\verb|crit|), are defined for QoS specification of a message, as illustrated in Listing \ref{lst:qos}.

\lstset{language=XML, caption=QoS specification in Etherware Message, label=lst:qos}
\begin{center}
\begin{lstlisting}
<EtherMsg type=... rel=... >
  <profile name=... ></profile>
  <content> ... </content>
  <ts value=... ></ts>
  <QoS crit=...  period=...  deadline=... wcet=...>
  </QoS>
</EtherMsg>
\end{lstlisting}
\end{center}

\subsubsection{Priority and Concurrent Processing}
Concurrency is a fundamental feature of any real system. Multiple tasks may be released concurrently, and a real-time scheduling policy prioritizes the execution order based on some rules, while satisfying their timing constraints. Thus, concurrency and priority are two key aspects of any real-time system.

\begin{figure}
\begin{center}
  \includegraphics[width=9cm]{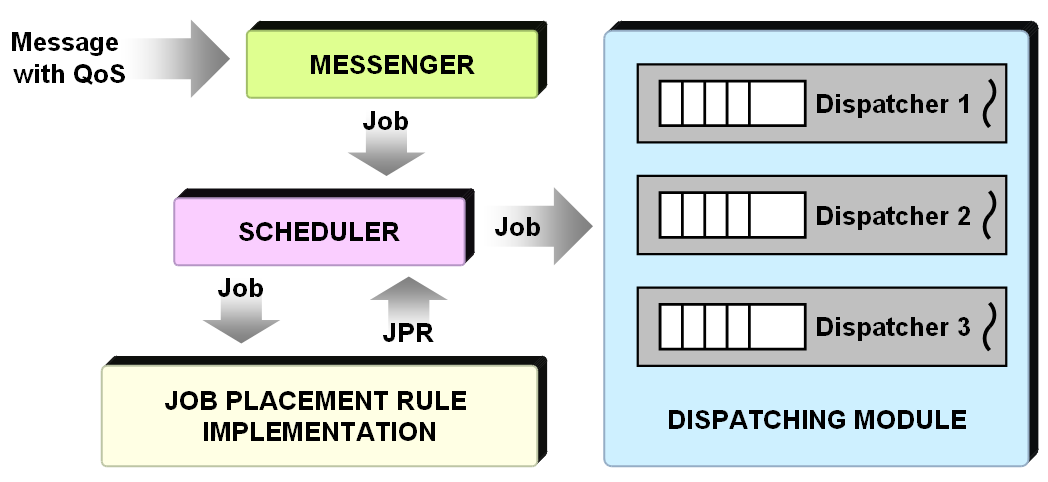}\\
  \caption{Real-time scheduling mechanism with three Dispatchers}\label{fig:schedule}
\end{center}
\end{figure}

As explained in Section \ref{sec:ether:arch}, Dispatcher is a software module inside the Etherware Kernel, for processing a \emph{job} (a scheduling entity in Etherware Kernel). For concurrent message processing, multiple Dispatchers can be used to form a \emph{dispatching module} as shown in Fig. \ref{fig:schedule}. For real-time message processing, Etherware uses a hierarchical prioritization mechanism \cite{gill:99}. First, each Dispatcher in a dispatching module is assigned a fixed priority\footnote{The specific priority set is given by the underlying software platform.} so that each message enqueued in a Dispatcher is processed at the fixed priority. The specific number of Dispatchers in a dispatching module and their corresponding priority levels are determined by a user-provided specification, called a \emph{Thread Scheduling Rule} (\emph{TSR}). Second, the job queue of a Dispatcher is a prioritized queue which orders jobs in the queue based on some information specified for a job. Owing to this hierarchical mechanism, Etherware can support various types of real-time scheduling policies, such as RM scheduling and non-preemptive EDF. An example of the implementation of a RM scheduling policy implementation is shown in Listing. \ref{lst:jpr}.

For predictable behavior, Etherware utilizes the priority-based scheduling mechanism of the underlying platform upon which Etherware is executed. This means that the execution order among Dispatchers is determined by the underlying operating system platform based on the fixed priorities assigned to each of them. Thus, Etherware Scheduler does not need to handle such priority-based scheduling. It is automatically taken care of by the platform. Instead, Etherware Scheduler performs the scheduling action at a higher level. Specifically, it determines the right place where a job should be put in the dispatching module. Due to the hierarchical nature of the prioritization in message processing, Etherware Scheduler uses two pieces of information, one to select a Dispatcher in dispatching module, and the other to find the right position in the job queue of the selected Dispatcher. This information is provided to Etherware Scheduler at runtime by a software module which implements a user defined scheduling rule, called the \emph{Job Placement Rule} (\emph{JPR}), which maps the QoS information specified in a message contained in a job to the position in the Etherware's dispatching module. Listing \ref{lst:jpr} is an example of the JPR implementation which utilizes the period information in a message to implement the RM scheduling policy. The corresponding experimental results are shown in Table \ref{tab:rms}.

\lstset{language=Java, caption=A pseudo-code example of JPR which implements the RM scheduling policy, label=lst:jpr}
\begin{center}
\begin{lstlisting}
/* QoS : period(P) in millisecond */
JPR queryJPR(Job job) {
 JPR jpr;
 if(job.P=80) jpr=(Disp#1, NULL);
 else if(job.P=200) jpr=(Disp#2, NULL);
 else if(job.P=350) jpr=(Disp#3, NULL);
 return jpr;
}
\end{lstlisting}
\end{center}

To test the real-time performance of Etherware, experiments were performed under several different conditions, depending on the execution time of a task with 200 (ms) period. In each of these different conditions, the activation periods and the execution times per activation of each task are measured to see how much these changes affect the execution behaviors of two other periodic tasks. As shown in Table \ref{tab:rms}, the task with the shortest period is not affected at all by the execution time changes of the task with the second shortest period. In contrast, the third task, which has the longest period, is affected a lot in its execution behavior by these changes. Thus, this result shows that the periodic tasks are in fact correctly scheduled under the RM scheduling policy.

\renewcommand{\arraystretch}{1.5}
\begin{table}
\begin{center}
\caption{Experimental results of task execution under RM scheduling policy } \label{tab:rms}
\begin{tabular}{|c|c|c|c|c|c|c|}
  \hline
  Test & \multicolumn{3}{|c|}{(mean, jitter) of Execution Time (ms)} & \multicolumn{3}{|c|}{(mean, jitter) of Period (ms)}  \\ \cline{2-7}
  Case & Task(80ms) & Task(200ms) & Task(350ms) & Task(80ms) & Task(200ms) & Task(350ms) \\ \cline{1-7}
  1 & (14.5, 1.1) & (42.4, 15.8) & (49.0, 51.2) & (80.9, 1.1) & (200.9, 29.5) & (350.9, 96.9) \\ \cline{1-7}
  2 & (14.5, 0.4) & (87.6, 16.0) & (56.3, 102.0) & (80.9, 1.2) & (200.9, 29.6) & (350.9, 156.1) \\ \cline{1-7}
  3 & (14.5, 0.3) & (129.9, 16.5) & (64.7, 153.0) & (80.9, 1.2) & (200.9, 29.5) & (350.8, 192.0) \\ \cline{1-7}
  4 & (14.5, 0.3) & (175.6, 17.6) & (350.8, 191.9) & (80.9, 2.0) & (200.9, 29.3) & (350.4, 191.9) \\ \cline{1-7}
  \hline
\end{tabular}
\end{center}
\end{table}

\section{Reliability for Networked Control Systems} \label{sec:ft}
\subsection{Fundamentals of Reliable System} \label{sec:ft:fundamental}
The reliability of a system is usually defined as the probability that a system will perform its functionality correctly throughout a duration of time. A technical measure of the reliability is the mean time between failures (MTBF), which is the sum of the mean time to failure (MTTF) and the mean time to repair (MTTR). The MTTF is a measure of how long a system is expected to operate correctly before a failure occurs, while the MTTR measures the difficulty of recovering a system after its failure. As the definition indicates, the failure of a system is at the heart of the discussion of the reliability. Therefore, we first introduce the definition of failure and two other fundamental concepts, \emph{error} and \emph{fault}.

\subsubsection{Fault, Error, and Failure}
Typically, all details about what are the acceptable behaviors of a system are described in a system specification. If the behavior of the system deviates from the specified acceptable behaviors, then this is called \emph{failure} of the system. The immediate cause of a failure is called an error. An \emph{error} is a part of an erroneous state, a system state which could lead to a failure by a sequence of valid state transitions. Finally, a \emph{fault} is defined to be the cause of an error. Thus, a fault is the root cause of a failure.

Depending on the view point, faults can be classified in many different ways \cite{selic:04}. A fault can be either \emph{transient} or \emph{permanent} based on the time duration that a fault can exist. Typically, a transient fault is much more problematic than a permanent fault since it is usually harder to diagnose. A fault can also be classified as a \emph{design fault} or \emph{operational fault} based on the underlying cause of the fault. The other classifications of faults are based on the symptoms caused by the fault, e.g., \emph{crash faults}, \emph{timing faults}, \emph{omission faults}, and \emph{Byzantine faults}. A crash fault is a fault which causes a system to crash so that it can never return to a valid operational state. When a system experiences an omission fault, it fails to perform its designated service even though it is still operating. Under a timing fault, a service provided by a system can be delayed. Lastly, a system behaves unpredictably with Byzantine faults. There is no specific patterns of symptoms caused by Byzantine faults.

\subsubsection{Fault Prevention}
One approach for achieving reliability of a system is to prevent system failures by ensuring that all possible causes of unreliability have been removed from the system before the system is deployed for its operation \cite{anderson:81}. This is called \emph{fault prevention}. As a first step toward fault prevention, a system needs to be carefully developed so that all faults which can be anticipated are removed during the development process. Various techniques or methodologies from software engineering, or formal methods, can be used during this stage to attempt to avoid introducing any faults into the system design.
However, it is not usually possible to guarantee that a developed system is completely free from faults. No matter how thorough the development process, there may always exist faults. The faults in the constructed system are therefore attempted to be removed through some experimental validation process. The implemented system is tested under various operating conditions to expose any existing faults, so that they can be removed. In some cases, artificial faults are introduced into the implemented system. This technique is called \emph{software fault injection} (\emph{SFI}) \cite{voas:98}. SFI tries to determine what could happen when faults are activated. The information collected through the SFI process can be used to both improve reliability of the system and also to estimate the resilience of the implemented system to faults.

\subsubsection{Fault Tolerance}
In general, the application of fault prevention techniques to a system is not sufficient to achieve high reliability. Given this fact, it is usually required that a system be fault tolerant in order to provide reliability despite the presence of faults. \emph{Fault tolerance} is the ability of a system to perform its function correctly even in the presence of internal faults \cite{selic:04}. There are four distinct activities which provide the general means by which fault tolerance can be implemented \cite{anderson:81}. These four activities constitute the basic principles which underly all fault tolerant systems.
Toward fault tolerance, errors caused by faults have to be detected first. Thus, \emph{error detection} is the first step for fault tolerance. The common techniques for error detection are replication checks, timing checks, coding checks, and so on (see \cite{anderson:81} for details). Once an error is detected, it may be necessary to asses the extent to which the system state has been damaged by the fault which manifested the detected error. This is known as \emph{damage confinement}. The next step is \emph{error recovery} which recovers the system from the erroneous system state to a valid error-free state. In the error recovery phase, there are basically two different approaches to recover the system state, backward error recovery and forward error recovery. In backward error recovery, the system state is restored to a past valid state which was checkpointed during normal operations. On the other hand, forward error recovery techniques drive the system to a new valid state which is produced by manipulating some portion of the erroneous current state. Finally, the forth step of fault tolerance is \emph{fault treatment}. To prevent reoccurrence of the error, it is necessary to identify and treat the fault. One issue in this procedure is that it may be hard or take a long time to identify faults which caused the errors, since the relationship between a fault and the corresponding errors is typically very complex.

\subsubsection{Software Fault Tolerance}
\begin{figure}
\begin{center}
  \includegraphics[width=11cm]{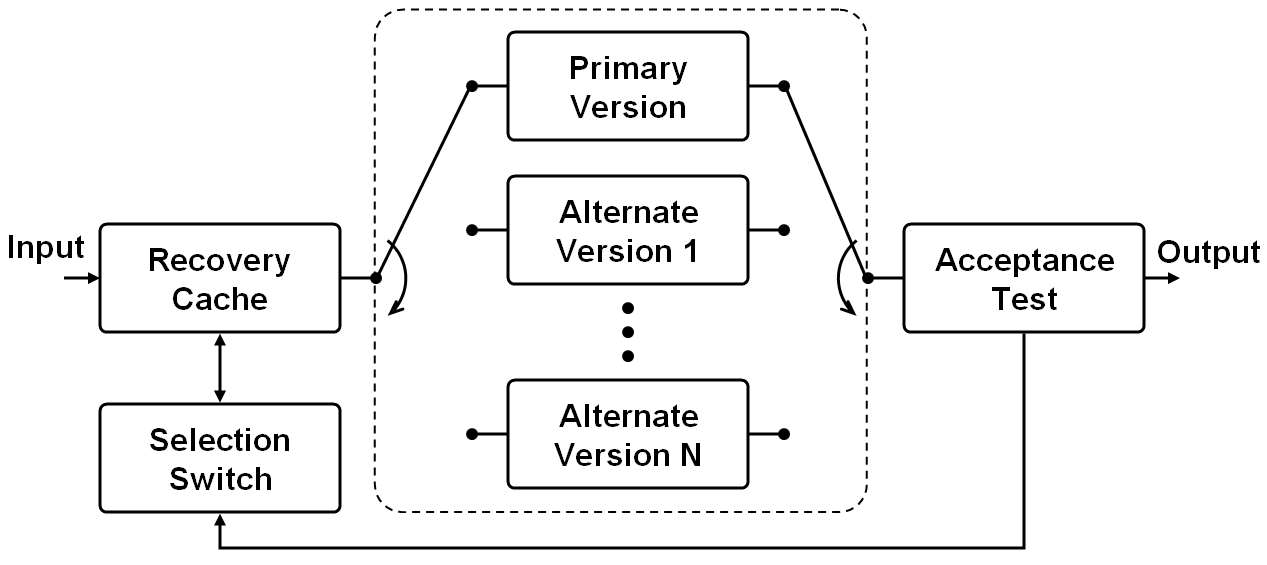}\\
  \caption{Recovery block scheme}\label{fig:ft:recovery}
\end{center}
\end{figure}

In this section, we introduce two main techniques for software fault tolerance, \emph{recovery block scheme} and \emph{N-version programming}. Basically, both the techniques depend on the effective utilization of redundancy with the expectation that components built differently should fail differently \cite{torres:00}. The basic configuration of the recovery block scheme is illustrated in Fig. \ref{fig:ft:recovery}. The first step in the recovery block scheme is checkpointing the current valid system state before executing any modules. Then the primary module is entered. Once the primary module completes its execution, the execution result of the primary module is tested by the acceptance test module to detect any error from the primary module if there is any. If the result is valid, then it is propagated as the output of the block. If any error is detected, then the error recovery process restores the primary module with the checkpointed state. Following the recovery, the same sequence of executions is repeated, except that the next module is used in place of the module that failed. If all of the modules fail, then it is considered as the failure of the recover block. One issue with the recovery block scheme is that the acceptance test module is highly application dependent \cite{torres:00}. Hence, its error detection logic is usually required to be implemented by the module developer.

\begin{figure}
\begin{center}
  \includegraphics[width=9cm]{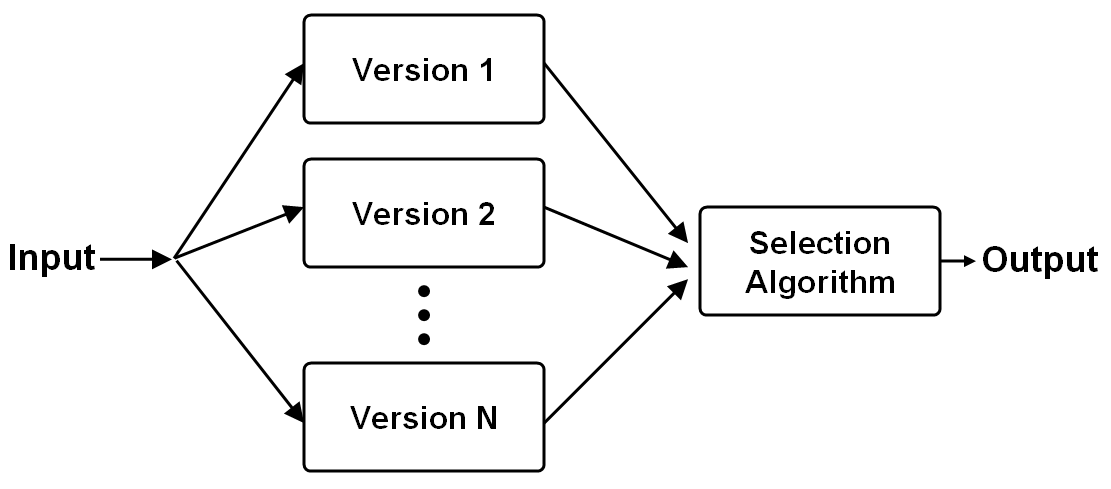}\\
  \caption{N-version programming model}\label{fig:ft:nvp}
\end{center}
\end{figure}

While the recovery block scheme requires an application dependent acceptance test module, the N-version programming model can use a generic decision algorithm to select the correct output \cite{torres:00}. Usually, voting algorithms such as Formalized Majority Voter, Generalized Media Voter, Formalized Plurality Voter, and Weighted Averaging Techniques, can be used as the generic decision algorithms in the design of the selection module (see \cite{lorczak:89} for details). Another significant difference between N-version programming and the recovery block is that the output is determined through a voting process. In a voting process, each of the modules processes the input first, and then their execution results are collected by the selection module. The selection algorithm determines the output based on the outputs from all modules and some decision rules. This procedure is illustrated in Fig. \ref{fig:ft:nvp}.

\subsection{Reliability Support in Etherware} \label{sec:ft:ether}
In this section, we describe how Etherware supports the reliability requirement of middleware for networked control systems.

\subsubsection{Local Temporal Autonomy}
\begin{figure}
\begin{center}
  \includegraphics[width=9cm]{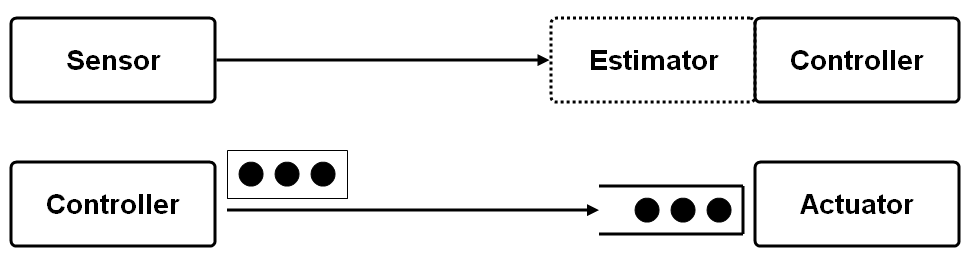}\\
  \caption{The LTA-based design principles}\label{fig:lta}
\end{center}
\end{figure}

\emph{Local temporal autonomy} (\emph{LTA}) is defined as the ability to operate correctly for a while in the presence of a failure of the other system components \cite{graham:04}. To accommodate the failure of a network connection, nodes or components in a networked control system, several design principles based on the LTA were proposed in \cite{graham:04,robinson:05}. The basic idea of these design principles is to reduce the inter-dependency between the interacting components. These principles can be easily understood by considering specific examples. As shown in the top figure in Fig. \ref{fig:lta}, in a networked control system, a controller relies on both the information from sensors and the communication network through which the sensor information is delivered. If one of these fails, a controller also fails. The proposed design principle suggests that an estimator be used, which is collocated, with the controller to estimate the sensor data. Then, any transient failure of either the sensor or communication network becomes transparent to the controller.

Another example is the case when a remote controller sends its computed control value to an actuator component which is collocated on the target plant and delivers the control to the plant. In this situation illustrated in the bottom figure in Fig. \ref{fig:lta}, the proposed design principle first employs a controller to compute multiple steps of future control values, and then sends them en bloc to the actuator. Secondly, an actuator is equipped with a buffer to hold the block of control values delivered from a controller. Clearly, this block computation and buffer mechanism can reduce dependence between a controller and an actuator so that any transient failure of either a controller or the communication network is transparent to the actuator. Etherware provides support for both these strategies \cite{robinson:05}.

\subsubsection{Component Model for Fault-Tolerance}
\begin{figure}
\begin{center}
  \includegraphics[width=9cm]{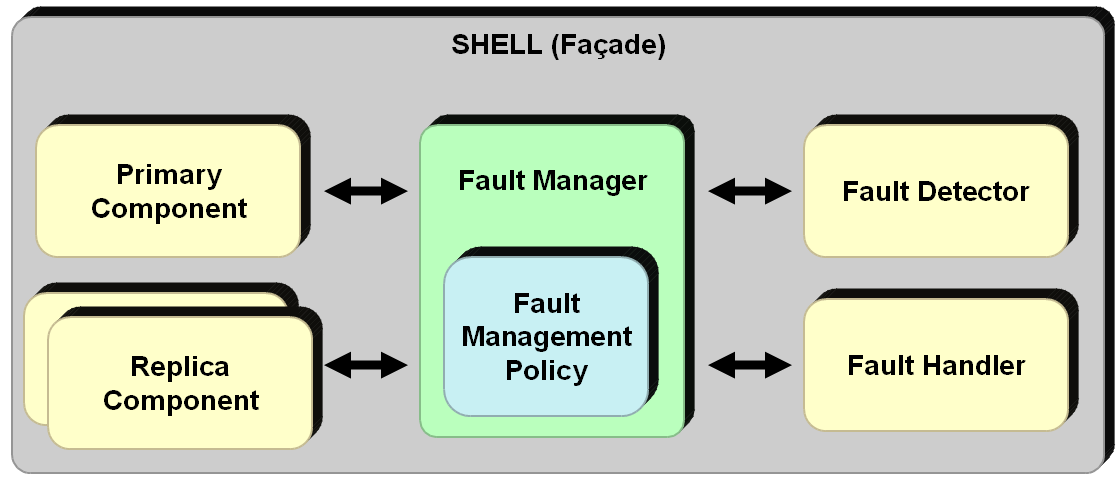}\\
  \caption{Fault-tolerant component model}\label{fig:ftcm}
\end{center}
\end{figure}

In addition to the LTA-based design principles, a \emph{fault-tolerant component model} (\emph{FT component model}) was proposed to support the reliability requirements in \cite{kdkim:08}. Basically, the FT component model is an extension of the original component model of Etherware. As shown in Fig. \ref{fig:ftcm}, it is designed to achieve both redundancy based fault-tolerance, and systematic fault detection and management. The redundancy based fault-tolerance is achieved by allowing multiple components, the primary and other replica components, to be executed within a Shell. The systematic fault detection and management is possible through the fault detector and fault handler elements in a Shell. Among the elements encapsulated in a Shell, the fault manager plays the central role in the fault tolerant operation. It coordinates all the interactions among elements within the FT component model. The \emph{fault management policy} (\emph{FM policy}) provides decisions about how to coordinate them. Depending on the FM policy, the FT component model itself can behave similar to either the recovery block mechanism or the N-version programming mechanism as explained in Section \ref{sec:ft:fundamental} (see \cite{kdkim:08} for details about the FM policy).

The fault detector can be used to detect a design fault related to a component's operational semantics. As an example, if the value computed by a component is typically expected to be both within some range and smooth, then the fault detector checks if the result from a component satisfies these conditions. If not, it reports this as a design fault to the fault manager, which in turn calls the fault handler to handle this fault. If there is a crash fault or a timing fault from other Etherware components, then these faults can also be managed by the fault handler. Unlike the design fault, detecting these faults cannot be done within a Shell since it usually requires a time-based delay detection mechanism such as a watchdog timer. Therefore, an additional Etherware service, called \emph{interaction fault detection service} (\emph{IFDS}), was proposed, and is being developed to detect interaction delays caused by another component's crash or timing fault.

\section{Case Study: Networked Inverted Pendulum Control System} \label{sec:dipcs}
In this section, to exemplify the usefulness of a middleware framework, specifically Etherware in this case study, in implementing a networked control system, we present the experimental results of using Etherware to control an inverted pendulum control system. Basically, two conclusions can be made which are supported by this case study. First, it can be easy to both develop and manage a networked control system with proper support from a middleware. Second, it is important to consider the non-functional requirements such as real-time and reliability, for correct operation of a networked control system.

Next, we briefly introduce the inverted pendulum control system which is used in the experiment.

\subsection{Inverted Pendulum Control System} \label{sec:dipcs:ipcs}
\begin{figure}
\begin{center}
  \includegraphics[width=12cm]{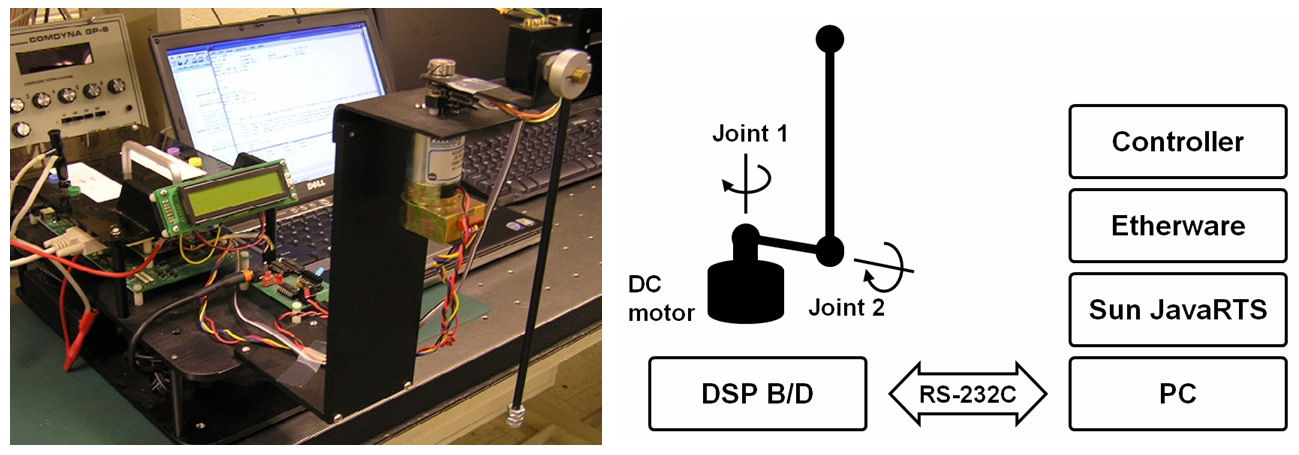}\\
  \caption{Inverted pendulum control system}\label{fig:ipcs}
\end{center}
\end{figure}

Fig. \ref{fig:ipcs} shows the inverted-pendulum system that is used in our experiment. As shown in the figure, it has two links. The link in the base is an active one which is actuated by a DC motor attached to it, while the other one is a passive link. The inverted pendulum system itself is equipped with a DSP board to measure the joint angles of both links and to apply the PWM signal to a DC motor. In our experiment, a controller, which implements a simple state feedback control law, is developed as an Etherware component, and it runs in a laptop which is attached to the inverted pendulum system through an RS-232C serial communication channel. For real-time operation of Etherware, we use the Sun JavaRTS\footnote{Sun JavaRTS is a real-time Java virtual machine which implements the real-time Java specification (RTSJ). For details about Sun JavaRTS and RTSJ, we refer the reader to \cite{javarts,rtsj:00}.} as the underlying platform of Etherware\footnote{Note that current Etherware is developed with Java programming language. Therefore, Etherware requires a Java virtual machine which provides the fixed-priority based scheduling mechanism for correct operation of Etherware's real-time Scheduler.}.

With support from Etherware's notification mechanism, explained in Section \ref{sec:ether:svcs}, the controller component is periodically activated to output its control action. In our experiment, the period is set to 15ms. In each period, the controller begins its execution by requesting the angle data from the DSP. Once it receives the measured angle data, it then computes a control output value and sends it back to the DSP so that the control command can be applied to control the inverted pendulum.

Thus, a periodic control action requires multiple interactions between a controller and the inverted pendulum system through the RS-232C communication network. Therefore, in addition to the predictable activation of a controller component in Etherware, the predictability and the reliability of the RS-232C communication network are also essential factors affecting the success of the inverted pendulum system. As explained in Section \ref{sec:rt:ether}, the predictability of the periodic activation of a controller component is guaranteed by the real-time scheduling mechanisms of Etherware. However, the predictability and the reliability of the RS-232C communication network is not guaranteed by the underlying platform in our implementation. Therefore, we adopted the state estimator LTA design principle to tolerate occasional communication errors, so as to achieve better periodic performance over the unreliable communication layer.

\begin{figure}
\begin{center}
  \includegraphics[width=7cm]{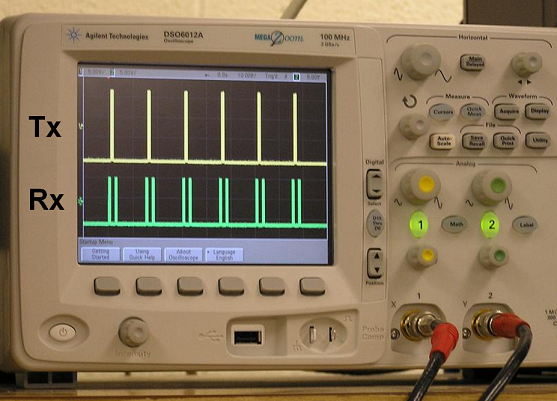}\\
  \caption{Periodic sensing and control action over RS-232C communication}\label{fig:rs232}
\end{center}
\end{figure}

Fig. \ref{fig:rs232} is a still oscilloscope image which captures the serial communication between the DSP and PC on which the Etherware-based controller is running. From the figure, we can observe the periodic interaction between the Etherware-based controller and the DSP. The upper signal in the scope image is the signal for feedback of angle data from DSP to PC, and the lower one is from PC to DSP for requesting angle data and sending a control command.

\subsection{Periodic Control under Stress} \label{sec:dipcs:control}
\begin{figure}
\begin{center}
  \includegraphics[width=8cm]{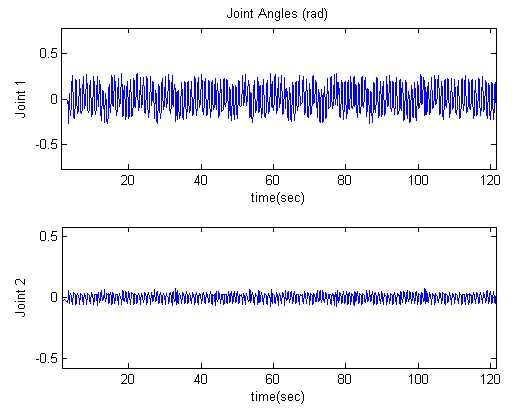}\\
  \caption{Periodic control of an inverted pendulum under stress}\label{fig:stress}
\end{center}
\end{figure}

The real-time performance of Etherware is verified through a stress condition that is imposed alongside the periodic control of inverted pendulum. To stress the computer on which a periodic control task is running, an extra computational task is made to run concurrently with the control task. In this experiment, the stress task is also executed periodically with a period of 5 seconds. Once it begins its execution, the stress task requires about 1 second to finish its computation. Considering that the period of control task is only 15 ms, as explained in the above section, 1 second is long enough to disturb the stability of the inverted pendulum, if the real-time performance had not been supported by Etherware. To achieve timely execution behavior of the periodic control task, the periodic notification message from Notifier to the periodic control task is specified to have a higher priority than that of the stress task. Fig. \ref{fig:stress} shows the experimental result of this experiment. In the experiment, the stress task starts its execution around 20 seconds after the system starts. As shown in the result, there is no apparent adverse affect on control performance even under the stress condition. This demonstrate Etherware's real-time performance.

\subsection{Runtime System Management} \label{sec:dipcs:manage}
As explained in Section \ref{sec:ether}, Etherware provides mechanisms which support the continuous evolution of a system after its deployment. The component model is at the heart of such mechanisms. \emph{The combination of these mechanisms for flexibility, and the real-time mechanisms for temporal predictability, generate capabilities which enable us to do runtime management of a system, especially a time-critical control system, without sacrificing the system's stability}. The specific capabilities that we aim to provide are \emph{controller upgrade} and \emph{controller migration}. In this section, we demonstrate these capabilities of Etherware.

\subsubsection{Controller Upgrade}
In this experiment, we show Etherware's capability for \emph{runtime component upgrade}. More specifically, a running inverted pendulum controller is replaced with a new controller which has better control performance, while still maintaining the stability of the inverted pendulum. To perform this upgrading process correctly without violating the stability of running system, it is important to externalize the state of the running controller and recover the state with a new controller timely. Etherware supports this operation though its component model which enables component upgrade and its real-time scheduling mechanism.

\begin{figure}
\begin{center}
  \includegraphics[width=12cm]{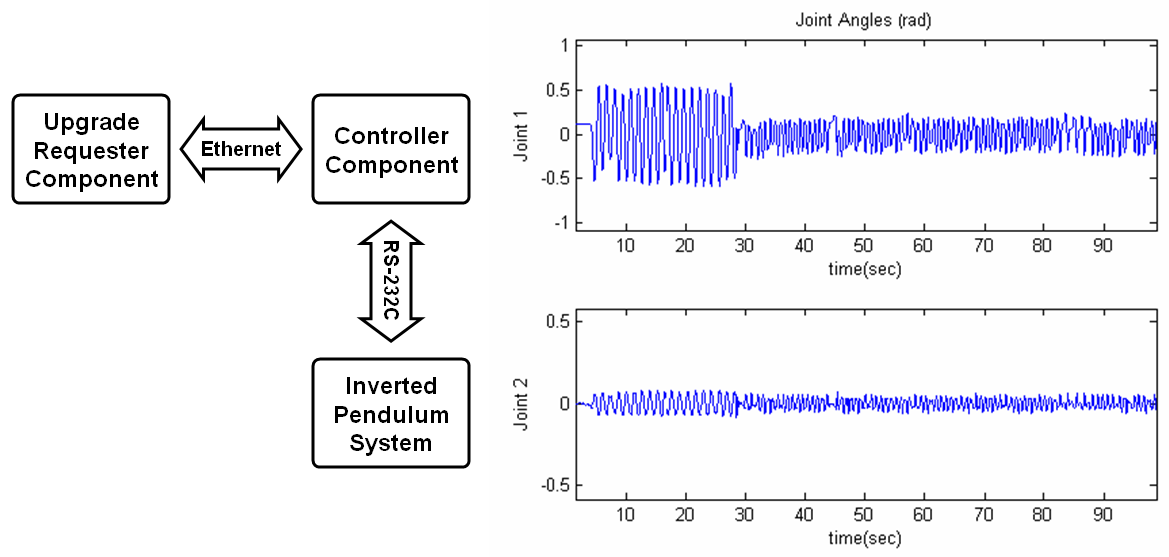}\\
  \caption{Joint angles of the inverted pendulum under runtime controller upgrade}\label{fig:upgrade}
\end{center}
\end{figure}

Fig. \ref{fig:upgrade} shows the configuration of an application for this runtime controller upgrade experiment. The periodic controller is running on a computer which is directly connected to the inverted pendulum system through a serial port. On the other computer, a component which requests the controller upgrade is running. In the experiment, the requester component sends a request message for better control performance around 30 seconds after the system starts. As shown in the figure, the control performance is improved around 30 seconds, at the time when the controller upgrade is requested. Thus, this result demonstrates Etherware's capability of real-time component upgrade.

\subsubsection{Controller Migration}
In this experiment, we demonstrate Etherware's capability for \emph{runtime component migration}. More specifically, a controller which controls the inverted pendulum is migrated from one computer to another at runtime while preserving the stability of the inverted pendulum. This type of capability can be very useful in a wide range of applications in optimizing the behavior of control systems. For example, if the network causes long delay from a specific computer in the network, then one may want to relocate the controller logic, i.e., component, to another computer which has less delay. For such runtime migration, several more steps of actions have to be performed, in addition to the runtime state externalization, as explained in previous section. Specifically, the externalized state itself has to be migrated to the destination where the controller will be migrated. Also, once a new instance of the controller is created at the destination node, the migrated state has to be recovered with the controller. Thus, component migration is a much more complex task than upgrade, in general. Furthermore, all these actions have to be performed in a timely manner so as to preserve the stability of the inverted pendulum. Therefore, Etherware's mechanisms for flexibility and predictability are essential to perform this runtime management operation. Moreover, Etherware's facilities make quite simple the development and deployment of such advanced capabilities.

\begin{figure}
\begin{center}
  \includegraphics[width=12cm]{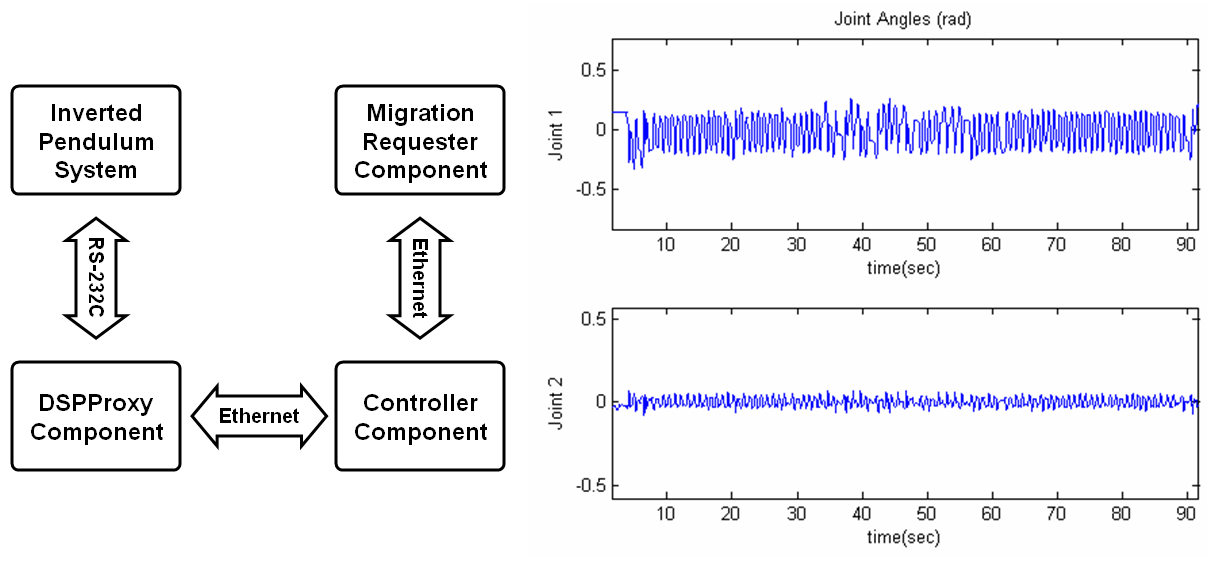}\\
  \caption{Joint angles of the inverted pendulum under runtime controller migration}\label{fig:migration}
\end{center}
\end{figure}

Fig. \ref{fig:migration} illustrates the application configuration for the runtime controller migration experiment. Initially, the inverted pendulum is controlled over the network by a controller which runs at remote computer. At the computer which is directly connected to the inverted pendulum, a component, called DSPProxy, is running to mediate the interaction between the controller and the DSP in the inverted pendulum system. At a third computer, another component which requests the controller migration process is running. In this experiment, the requester component sends out a request message, which requests the migration of the controller from its current location to the computer which has the direct connection to the inverted pendulum. This request is made around 40 seconds after the system starts. Then Etherware performs the controller migration process. As shown in Fig. \ref{fig:migration}, the stability is maintained even while the controller is migrating to a new computer node. Thus, this result demonstrates Etherware's capability of real-time component migration.

\section{Conclusion}
In this paper, we have discussed fundamental characteristics which are common to any networked control systems. The four characteristics, which are large-scaleness, openness, time-criticality, and safety-criticality, are identified in Section \ref{sec:dcs:char}. Due to these fundamental characteristics, there is a need for a middleware, which enables us to realize such complex control systems. Indeed our thesis is that such a middleware is important for the future of networked control systems.

As an underlying platform for networked control systems, a middleware is required to satisfy domain requirements which are necessitated by the domain characteristics. There are several fundamental functionalities that have to be provided by any middleware to satisfy both the operational and management domain requirements which are induced by the characteristics of a distributed system. Etherware, a middleware developed at the University of Illinois, is an example of such a middleware for networked control system.

In addition to the functional domain requirements such as operational and management requirements, a middleware framework is typically expected to also support the non-functional domain requirements such as timeliness and reliability. We have presented Etherware's approach to support the timeliness requirements. We have also addressed the issue of faults and approaches toward fault prevention and tolerance, Etherware's LTA-based design principles, and the fault-tolerant component model.

The performance of a networked inverted pendulum control system is provided as a case study of a middleware based networked control system. In the presented system, Etherware is used as a middleware framework which for rapid and evolvable control application development. We have exhibited complex and important run time capabilities such as controller migration and controller update, to highlight the sophisticated capabilities that a middleware such as Etherware can provide. We have thus experimentally demonstrated Etherware's flexibility and temporal predictability properties.

\begin{acknowledgement}
This material is based upon work partially supported by AFOSR under Contract No. FA9550-09-0121, ARO under Contract No. W911NF-08-1-0238, and NSF under Contract Nos. NSF ECCS-0701604, CNS-07-21992, and CCR-0325716.
\end{acknowledgement}

%


\nocite{baliga:05,bollella:00,buttazzo:04,GoF:94,gill:99,graham:04,kdkim:08,tan:07,java:05,jvm:99,rtsj:00,lund:98,javarts,
metro:03,gill:99,corba,graham:09,vinoski:04,puder:05,stankovic:88,liuland:73,sha:04,sha:90,selic:04,anderson:81,torres:00,
lorczak:89,convergencelab,xml:w3c,robinson:05}
\bibliographystyle{splncs}
\bibliography{NCSBOOK}
\end{document}